\documentclass[aps, prl, showpacs, twocolumn]{revtex4}
\usepackage{graphics}
\usepackage{graphicx}
\usepackage{amssymb}
\usepackage{amsmath}
\usepackage{bm}

\newcommand{\bra}[1]{\langle #1 \vert}
\newcommand{\ket}[1]{\vert #1 \rangle}

\begin{document}

\title{Holographic quantum computing}
\author{Karl Tordrup}
\author{Antonio Negretti}
\author{Klaus M\o{}lmer}
\affiliation{Lundbeck Foundation Theoretical Center for
Quantum System Research, Department of Physics and Astronomy,
University of Aarhus, DK-8000 Aarhus C, Denmark}

\date{\today}

\begin{abstract}
We propose to use a single mesoscopic ensemble of trapped polar molecules for quantum computing. A "holographic quantum register" with hundreds of qubits is encoded in collective excitations with definite spatial
phase variations. Each phase pattern is uniquely addressed by optical Raman processes with classical
optical fields, while one- and two-qubit gates and qubit read-out are accomplished by transferring the qubit states to a stripline microwave cavity field and a Cooper pair box
where controllable two-level unitary dynamics and detection is governed by classical microwave fields.
\end{abstract}


\maketitle

In classical computer science holographic data storage is poised to provide the next generation in
digital media\cite{holoMemNature,holoMemScience}. The defining characteristic of this method is that
information is stored globally rather than on specific sites in a storage medium. Current
investigations of quantum memory components include similar ideas for storage of optical information
in ensembles of atoms\cite{atomMemPolzik,atomMemKimble,Hau,Kuzmich} and molecules\cite{crystalMemory}.
\begin{figure}
\includegraphics[width=8cm]{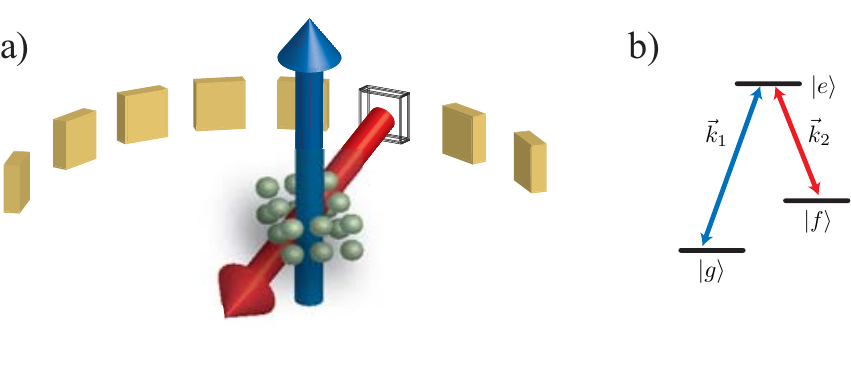}
\caption{(Color online). (a) By varying the direction of a
control field $\Omega_2(t)e^{i\vec{k}_2\cdot\vec{x}}$, an
incident single photon with wave vector $\vec{k}_1$ may be
transferred to different collective storage modes with wave
vector $\vec{q}=\vec{k}_1-\vec{k}_2$. (b) The levels
$\ket{g}$ and $\ket{f}$ are coupled by a two photon process
leaving no population in the electronically excited state
$\ket{e}$.}\label{fig:holo}
\end{figure}

In the quantum version of holographic storage one can
envisage $N$ atoms or molecules in a lattice initially all
populating the same internal quantum state $\ket{g}$ [see
Fig.\ \ref{fig:holo}(b)]. The quantum information in an
incident weak field $\Omega_1e^{i\vec{k}_1\cdot\vec{x}}$ is,
by the assistance of a control field
$\Omega_2(t)e^{i\vec{k}_2\cdot\vec{x}}$ and the Hamiltonian
\begin{equation}\label{eq:Hk}
H=\sum_{j=1}^N
\Omega_1e^{i\vec{k}_1\cdot\vec{x}_j}\ket{e_j}\bra{g_j} +
\Omega_2e^{i\vec{k}_2\cdot\vec{x}_j}\ket{e_j}\bra{f_j} +
\text{h.c.},
\end{equation}
transferred onto a collective matter-light excitation. The Hamiltonian has a dark state, which maps
the field into a collective population of the state $\ket{f}$ by turning off $\Omega_2(t)$. In order
for this storage mechanism to work, the optical depth of the sample must be large\cite{polariton},
which may indeed be the case for a sufficiently large sample of atoms or molecules.

The coupling in Eq.\ (\ref{eq:Hk}) can be used to map a
single-photon state to the collective phase pattern state
$\ket{f,\vec{q}} \equiv 1/\sqrt{N} \sum_j e^{i\vec{q}\cdot
\vec{x}_j} \ket{g_1 \ldots f_j \ldots g_N}$, where
$\vec{q}=\vec{k}_1-\vec{k}_2$ is the wave number difference
of the two fields. For an extended ensemble with a large
number $N$ of atoms or molecules, phase pattern states with
sufficiently different wave numbers approximately fulfill the
orthogonality relation
\begin{equation}\label{eq:orthogonal}
\bra{f,\vec{q}_1}f,\vec{q}_2\rangle =
\frac{1}{N}\sum_{j=1}^Ne^{i(\vec{q}_2-\vec{q}_1)\cdot
\vec{x}_j} \approx \delta_{\vec{q}_1\vec{q}_2}.
\end{equation}
Such collective excitations can be used to simultaneously encode a large number of qubits by
associating the logical state $\ket{b_1 b_2 \ldots b_K}$ ($b_i=0,1$) with the collective state
$\prod_i(a_{\vec{q}_i}^{\dag})^{b_i}\ket{g_1 g_2\ldots g_N}$, where $a_{\vec{q}_i}^{\dag}=\sum_{j=1}^N
e^{i\vec{q}_i\cdot \vec{x}_j}\ket{f_j}\bra{g_j}$. That is, the identification of $K$ orthogonal [in
the sense of Eq. (\ref{eq:orthogonal})] wave vectors $\vec{q}_i$ allows construction of a $K$ qubit
register. Addressing different qubits is then merely a question of applying laser beams from different
directions to adhere to the phase matching condition, as illustrated in Fig.\ref{fig:holo}(a).

\begin{figure}
\includegraphics[width=6cm]{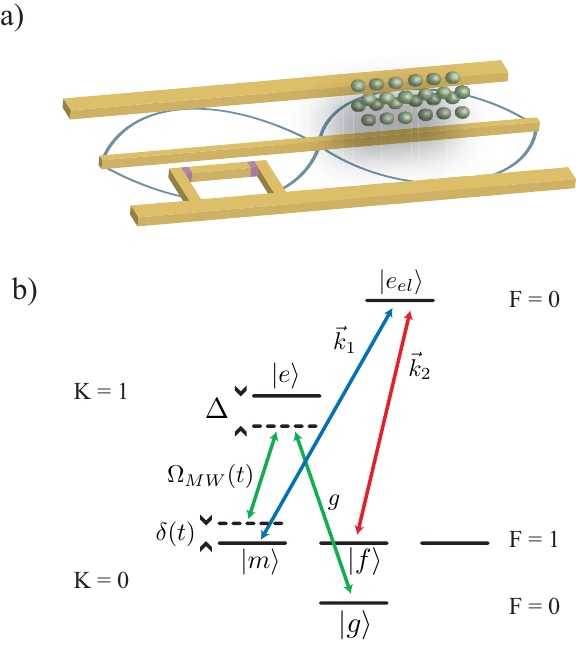}
\caption{(Color online). (a) A stripline cavity field is
coupled to a Cooper pair box and an ensemble of trapped polar
molecules. (b) Level structure of a molecule with a
$^2\Sigma_{1/2}$ electronic ground state and nuclear spin
$1/2$. Only levels in the rotational ground state ($K=0$) are
populated. The $F=0$ hyperfine ground state $\ket{g}$ is
coupled with a Raman process involving the cavity field and a
microwave source to the $\ket{F=1,m_F=+1}$ triplet state
$\ket{m}$ via a $K=1$ rotationally excited state. Two optical
fields provide further coupling to the triplet
$\ket{F=1,m_F=0}$ state $\ket{f}$ via an electronically
excited state. }\label{fig:setup}
\end{figure}

We shall present a proposal for a universal quantum computer, with a full register of qubits stored
and addressed in the way described above. The challenges lie in preventing multiple excitations
involving terms of the form $\prod_i(a_{\vec{q}_i}^{\dag})^{b_i}\ket{g_1 g_2\ldots g_N}$ with any $b_i
>1$, and in providing the interactions necessary to drive single-qubit and two-qubit gates. The
physical system we shall consider is a sample of cold polar molecules, trapped at an antinode of the
quantized electromagnetic field of a superconducting stripline resonator [see Fig.\ref{fig:setup}(a)]
\cite{hybrid}. The stripline cavity field is characterized by a wavelength in the
cm range, and by a transverse modal extent of only few $\mu$m \cite{hybrid,photonNo}. The associated
small mode volume implies a strong quantum field amplitude of even a single photon, and due to the
collective enhancement of the field-matter interaction, molecules confined to an elongated volume few
$\mu$m away from the axis of the waveguide are strongly coupled to the cavity field. A Cooper pair
box (CPB) is situated at an adjacent antinode of the field. The CPB consists of a superconducting
island onto which quantized charge may tunnel through insulating barriers \cite{Vion2002}. The large
dipole moment associated with this displacement of charge together with the large value of the single
photon electric field makes it possible to couple the field resonantly to the CPB with a
Rabi frequency much higher than the decay rates of the cavity field and of the CPB excitation
\cite{YaleCPB,Blais2004}.
The CPB has non-equidistant energy levels and when operated at cryogenic temperatures the
two lowest-lying states act as a two-level system, effectively controllable under
illumination by a classical resonant microwave field \cite{Vion2002}. We shall now show how this hybrid
architecture allows to perform the operations required for universal quantum computation on our
multi-qubit register by transferring the quantum state between the molecules and the CPB.

First, we shall adopt the molecular level scheme presented in Fig.\ref{fig:setup}(b). In addition to
hyperfine states $\ket{g}$ and $\ket{f}$ for qubit storage in the rotational ground state, we introduce
an auxiliary rotational ground state $\ket{m}$ and rotationally and electronically excited states
$|e\rangle$ and $|e_{el}\rangle$, respectively. As proposed in Ref.\cite{hybrid}, we
can exploit a classical microwave field $\Omega_{MW}$ and the collective enhancement of the
molecule-field interaction to resonantly transfer a single quantum of excitation from the cavity field
to a single collective excitation of the molecular ensemble via the rotationally excited state
$|e\rangle$. The cavity one-photon state is thus transferred to a collective molecular state
$\ket{m,\vec{0}} \equiv 1/\sqrt{N} \sum_j \ket{g_1 \ldots m_j \ldots g_N}$ with a single molecule
populating state $\ket{m}$ and with a vanishing phase variation across the ensemble (because it is
transferred from the ground state by means of long wavelength microwave fields). The state amplitude
in $\ket{m,\vec{0}}$ can subsequently be transferred to the collective state $\ket{f,\vec{q}_i}$ by a
STIRAP process with classical optical fields\cite{stirap}. The beam with wave vector $\vec{k}_1$ is kept fixed while the beam with wavenumber $k_2$ is rotated in a plane to give different $\vec{q}_i =
\vec{k}_1 - \vec{k}_{2,i}$. By inverting the order of the fields in the STIRAP process, the state
vector amplitude can be returned to $\ket{m,\vec{0}}$ at later times from which a transfer to the
cavity field is possible.

Before discussing the details of the various transfer processes and the coupling to the Cooper pair
box, let us outline how one stores multiple qubits in the same molecular ensemble. Imagine that a
single cavity qubit in the form of a superposition of zero and one photon states has been transferred
to the corresponding superposition of the collective molecular ground state and the state
$\ket{f,\vec{q}_1}$. We now wish to transfer a second photonic qubit to another wave vector pattern
state $\ket{f,\vec{q}_2}$. Using the collectively enhanced molecule-field coupling, we may transfer
the amplitude of the one-photon state to the $\ket{m,\vec{0}}$ state as above, but the subsequent
coupling of $\ket{m}$ and $\ket{f}$ with a wave vector $\vec{q}_2$ couples the already encoded
excitation in $\ket{f,\vec{q}_1}$ to $\ket{m,\vec{q}_1-\vec{q}_2}$. Furthermore since the STIRAP
process occurs in the "wrong order" for the latter coupling, the molecular excited state will also
become populated in the process. This problem is solved if we initially apply an inverted STIRAP pulse
with wave vector $\vec{q}_2$, so that the first qubit is reliably transferred to the intermediate
state $\ket{m,\vec{q}_1-\vec{q}_2}$ while the second qubit remains in the cavity. Then we transfer the
field excitation to the state $\ket{m,\vec{0}}$, and with a final STIRAP process with wave vector
$\vec{q}_2$, the two states are transferred to $\ket{f,\vec{q}_1}$ and $\ket{f,\vec{q}_2}$. Note that
the collective enhancement of the coupling is crucial for this protocol to work. When we map the
cavity state to the molecules, it is possible for the amplitude in the intermediate state
$\ket{m,\vec{q}_1-\vec{q}_2}$ to be converted into a field excitation in the cavity, but due to the
phase variation across the sample this coupling is suppressed, while the field coupling to the zero
wave vector state experiences the collective enhancement factor $\sqrt{N}$. To go beyond two qubits we
simply apply the same steps, such that both storage and retrieval of a molecular qubit in state
$\ket{f,\vec{q}_j}$ is preceded by shifting all $\ket{f,\vec{q}_i}$ qubit states "backwards" to
$\ket{m,\vec{q}_i -\vec{q}_j}$. The collective enhancement ensures that only $\ket{m,\vec{0}}$ can be
mapped to or from the cavity before all states are brought "forwards" back to $\ket{f,\vec{q}_i}$.

We now turn to the details of the physical proposal and the transfer processes. The interaction of the
CPB and the cavity is described by a Jaynes-Cummings type Hamiltonian
\begin{equation}
H_{\text{CPB}} = g_c\left(\sigma^-c^{\dag} + \sigma^+c\right)+\delta_{\text{CPB}}(t)\sigma^+\sigma^-,
\end{equation}
where $\sigma^+$ ($\sigma^-$) is the CPB raising (lowering) operator, $c^{\dag}$ ($c$) the cavity
field creation (annihilation) operator and $\delta_{\text{CPB}}(t) = \omega_{\text{CPB}}(t) -
\omega_c$ is the tunable CPB detuning with respect to the cavity. This coherent coupling has been
demonstrated in a number of experiments\cite{YaleCPB,photonNo}.

The cavity is coupled to the molecules through a Raman transition [see Fig.\ \ref{fig:setup}(b)]
involving the cavity field with coupling strength $g$ and a classical microwave field $\Omega_{MW}(t)$
which are both detuned by $\Delta$ from the rotationally excited state $|e\rangle$. We describe the
dynamics with the Hamiltonian
\begin{equation}\label{eq:Hmol}
H_{\text{M}} = g_{\text{eff}}(t)(c^{\dag} \sum_j |g_j\rangle\langle m_j|  + c \sum_j
|m_j\rangle\langle g_j|) - \delta(t)c^{\dag}c,
\end{equation}
where the effective coupling strength $g_{\text{eff}}(t) = \Omega_{MW}(t)g/2\Delta$ is
collectively enhanced by the square root of $N_0$, the number of molecules in the ground state.
$\delta (t)$ is the two-photon Raman detuning and $g\sim2\pi\times10-100$kHz\cite{cavityMolCoupling}.
If we consider the case of $N \sim 10^5 - 10^6$ molecules the collective enhancement increases the
light-matter coupling by three orders of magnitude compared to the addressing of individual molecules,
and if we encode no more than $K \sim$ a few hundred qubits, $N-K \leq N_0 \leq N$,  we can neglect
the depletion of the ground state and treat $N_0$ as a constant. Effective transfer frequencies are
in the range of $\sqrt{N_0}g_{\text{eff}}\sim  2\pi\times 1-10$ MHz \cite{hybrid}. As indicated previously we use the
state $\ket{m,\vec{0}}$ only as an intermediate station, and we use the optical STIRAP process via
an electronically excited state to connect to the final state $\ket{f,\vec{q}_i}$ in a few tens of nanoseconds\cite{Schiemann93}.

To perform a single qubit rotation, we map the qubit to the cavity and transfer it to the CPB. Single
qubit rotations can then be performed on the CPB using microwave pulses. Two-qubit gates can be
realized by transferring one qubit to the CPB and another to the cavity. We prevent their interaction
with a large initial detuning, $\delta_{\text{CPB}}(t=0)/g_c \gg 1$. To couple the CPB and the cavity
field, $\delta_{\text{CPB}}(t)$ is adiabatically tuned close to resonance and then back to
$\delta_{\text{CPB}}(T)/g_c \gg 1$, causing the combined CPB-cavity dressed states to acquire state
dependent phases. By choosing the form of $\delta_{\text{CPB}}(t)$ appropriately it is possible to
implement a fully entangling controlled phase gate\cite{hybrid,hybridMe}.
\begin{figure}
\includegraphics[width=8.6cm]{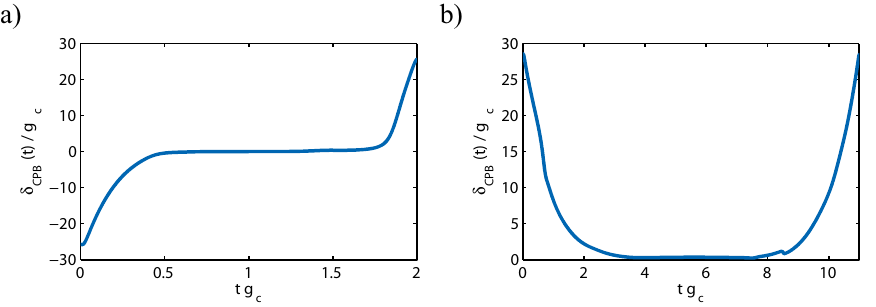}
\caption{(Color online). (a) Functional form of $\delta_{\text{CPB}}(t)$ implementing a SWAP between
the cavity and CPB. The coupling $g_c$ cannot be turned off so a sweep from negative to positive
detuning transferring a cavity state to the CPB must be followed by the time reversed sweep to
transfer the state back to the cavity.  (b) Functional form of $\delta_{\text{CPB}}(t)$ implementing a fully entangling conditional
phase gate between a CPB and a cavity qubit.}\label{fig:detunings}
\end{figure}
Using quantum optimal control theory\cite{OCT2} we have improved the SWAP and conditional phase shift
operations. The resulting optimal variations of $\delta_{\text{CPB}}(t)$, shown in Fig.\ref{fig:detunings},
achieve infidelities below the $10^{-4}$ level. Read out is carried out by transferring
a molecular qubit to the CPB.  This changes the resonant transmission properties of the cavity, and readout with a classical microwave pulse can be achieved in a few tens of nanoseconds\cite{Blais2004,Gambetta2008}.

The decoherence time of the CPB when operated at the so called "sweet spot" is determined mainly by
charge noise dephasing with $T_2 \sim 1$ $\mu$s \cite{Vion2002,Blais2004}. By replacing the
traditional Cooper pair box
with a transmon design\cite{transmonImplement} the main decoherence channel is spontaneous emission
relaxation with $T_1 \sim 4$ $\mu$s\cite{transmon08}. With feasible coupling strengths of up to $g_c
\sim 2\pi \times 200$MHz\cite{cpbQI} thousands of operations can be carried out within the coherence
time of the CPB. The stripline cavity can be manufactured with photon loss rates as low as $2\pi
\times 5$kHz\cite{cpbQI}. This corresponds to a decay probability during a SWAP or conditional phase
gate of the order $10^{-4}$ and standard error correcting codes may be employed against errors occuring in the CPB or cavity during gate operation\cite{Steane2003}.

In order to achieve long coherence times for the molecular ensemble qubits, it is advantageous to employ spin states for the
internal states $\ket{g}$, $\ket{m}$ and $\ket{f}$\cite{crystalMemory}. In a molecule such as CaF with
a $^2\Sigma_{1/2}$ ground state coupled by the hyperfine interaction to a nuclear spin of $I=1/2$ we
can use the singlet $F=0$ for $\ket{g}$ and the two $F=1$ triplet states with $m_F=1$ and $m_F=0$ for
$\ket{m}$ and $\ket{f}$ as illustrated in Fig.\ \ref{fig:setup}(b). Qubits encoded in different spin
states of the rotational ground state are protected from dipole-dipole interactions, and the
decoherence rate due to higher order spin-flip processes is below the $1$ Hz level\cite{crystalMemory}.
Similarly, typical dephasing rates of the hyperfine ground states are also below the $1$ Hz level. The
holographic storage is not prone to the conventional independent qubit errors, since single molecule
disturbances affect all qubits, but only very weakly \cite{errorCorrect}.

The trapping and cooling of a molecular sample with only $\mu$m separation to the superconducting
wire elements is a major challenge under current investigation in connection with a host of
interesting proposals involving molecules coupled to quantized fields\cite{hybrid, cavityMolCoupling}. Trapping of cold polar molecules in optical lattices was proposed and analyzed in\cite{DeMille2002}, but a clever design will have to be made to allign the optical field
with the cavity stripline. Electrostatic traps for molecules have been demonstrated \cite{Meek2008}, and
here further development is needed to miniaturize the design to the requested spatial dimension.

Since
the molecules are excited with different spatial phase factors, and the collectively stored quantum states
are read out according to these factors, they must not move around freely and the interaction with the
STIRAP laser fields must not excite motion of the individual molecules on the spatial scale set by the
wavelength of the phase patterns. These demands are significantly reduced if the STIRAP lasers are
made almost co-propagating, e.g., with one perpendicular to the trap axis, and the other one making an
angle just big enough for the resulting phase patterns to be orthogonal in the sense of
Eq.(\ref{eq:orthogonal}). For equidistant molecules along a trap axis of length $L$, this will be the
case for (small) angles of the incident field with respect to the normal, obeying
$k_2L\sin\theta_n=n\cdot 2\pi$, where $k_2$ is the wavenumber of the STIRAP beam with varying angle of
approach. With $L$ on the order of $5\text{mm}$ and $\lambda=2\pi/k_2$ on the order of $500\text{nm}$, a hundred
phase pattern states with $-50 \leq n \leq 50$ correspond to a narrow angular range of
$\vert\theta_n\vert \leq 0.3^\circ$, and to correspondingly low wave number excitations along and
perpendicular to the trap axis. In an optical lattice, the motion of every individual molecule is restricted
to less than the optical wave length, while in a larger electrostatic trap, the molecules may form self-organized
crystalline structures\cite{crystalMemory}, and also in that case their individual and collective motion may be kept 3-4 orders of magnitude smaller than
the shortest phase pattern wavelength of the order $L/50 = 100 \mu\text{m}$, so that the molecules are
addressed with the correct spatial phase factors by the STIRAP laser fields.

In conclusion we have described a holographic quantum information system able to support hundreds of
qubits and thousands of one- and two-qubit operations. If progress is made on the tight trapping of
regular structures of molecules, it is conceivable that even a thousand qubits may be implemented in a
single molecular sample. Alternatively, multiple hundred-qubit samples may be localized in different
anti-nodes of the cavity fields and hence reach scalability of the design. We reiterate, however, that
the coupling of a few molecules to the quantized cavity field is too small to be useful for reliable
transfer of quantum states, and it is essential to encode as many qubits as possible in as large
samples as possible to benefit from the much stronger coupling due to the collective enhancement.

This work was supported by the European Commission through
the Integrated Project FET/QIPC "SCALA".

\bibliography{minbib}

\end{document}